\documentclass[pteplogo,letterpaper]{ptephy_v1}
\Year{2024}
\pdfoutput=1
\usepackage[utf8]{inputenc}
\usepackage[colorlinks=true,urlcolor=blue,anchorcolor=blue,citecolor=blue,filecolor=blue,linkcolor=blue,menucolor=blue,linktocpage=true,bookmarksopen,pdfa=true,unicode]{hyperref}

\usepackage{mathtools}
\usepackage{lmodern}
\usepackage[T1]{fontenc}
\usepackage{bbm}
\DeclareMathAlphabet{\mathbfi}{OML}{cmm}{b}{it}
\usepackage{mathrsfs}

\let\originalleft\left
\let\originalright\right
\renewcommand{\left}{\mathopen{}\mathclose\bgroup\originalleft}
\renewcommand{\right}{\aftergroup\egroup\originalright}
\makeatletter
\newcommand{\biggg}{\bBigg@\thr@@}
\newcommand{\Biggg}{\bBigg@{3.5}}
\makeatother

\makeatletter

\def\@spliteq#1{\begin{equation}\begin{split}#1\end{split}\end{equation}}
\def\@spliteqstar#1{\begin{equation*}\begin{split}#1\end{split}\end{equation*}}
\def\splitequation{\collect@body\@spliteq}
\expandafter\def\csname splitequation*\endcsname{\collect@body\@spliteqstar}

\expandafter\def\csname endsplitequation*\endcsname{\ignorespacesafterend}
\makeatother

\makeatletter
\g@addto@macro\@floatboxreset\centering
\makeatother

\renewcommand{\vec}[1]{{\ifnum9<1#1\mathbf{#1}\else\ifcat\noexpand#1\relax\boldsymbol{#1}\else\mathbfi{#1}\fi\fi}}
\newcommand{\mathe}{\mathrm{e}}
\newcommand{\mathi}{\mathrm{i}}
\newcommand{\total}{\mathop{}\!\mathrm{d}}
\newcommand{\abs}[1]{{\left\lvert{#1}\right\rvert}}
\newcommand{\eqend}[1]{\,#1}
\newcommand{\bra}[1]{\left\langle{#1}\right\vert}
\newcommand{\ket}[1]{\left\vert{#1}\right\rangle}

\makeatletter
\newsavebox{\@brx}
\newcommand{\llangle}[1][]{\savebox{\@brx}{\(\m@th{#1\langle}\)}%
  \mathopen{\copy\@brx\kern-0.5\wd\@brx\usebox{\@brx}}}
\newcommand{\rrangle}[1][]{\savebox{\@brx}{\(\m@th{#1\rangle}\)}%
  \mathclose{\copy\@brx\kern-0.5\wd\@brx\usebox{\@brx}}}
\makeatother

\usepackage{tikz}
\protected\def\doublecone{\tikz[baseline=0.5pt]{\draw (0,4pt) arc(190:350:5pt and 1.5pt); \draw (0,4pt) -- (5pt,9pt) -- (10pt,4pt) -- (5pt,-1pt) -- cycle; \draw[densely dotted] (0,4pt) arc(170:10:5pt and 1.5pt);}}

\allowdisplaybreaks

\makeatletter
\renewcommand{\titlepagefootline}{%
  \fontsize{6.5}{8.5}\selectfont
  \vbox{\vspace*{12pt}
\hbox to 436.9pt{\hspace*{-1.9pt}\hspace*{-1.4pt}}
\hbox to \textwidth{$\copyright$ The Author(s) 2023. Published by Oxford University Press on behalf of the Physical Society of Japan.\hfill}
\hbox to \textwidth{This is an Open Access article distributed under the terms of the Creative Commons Attribution License\hfill}
\hbox to \textwidth{(http://creativecommons.org/licenses/by-nc/3.0), which permits unrestricted use,\hfill}
\hbox to \textwidth{distribution, and reproduction in any medium, provided the original work is properly cited.\hfill}
 \vspace*{12pt}}}
\makeatother

\newcommand{\aemail}[3][]{\affil[#2]{\if #1\empty\else#1 \fi E-Mail: \href{mailto:#3}{#3}}}

\begin{document}

\title{Entropy-area law and temperature of de Sitter horizons from modular theory}

\author[1,2,$\ast$]{Edoardo D'Angelo}
\author[3,$\dagger$]{Markus B. Fr\"ob}
\author[1,$\ddag$]{Stefano Galanda}
\author[4,5,$\S$]{Paolo Meda}
\author[3,$\P$]{Albert Much}
\author[6,$\parallel$]{Kyriakos Papadopoulos}

\affil[1]{Dipartimento di Matematica, Dipartimento di Eccellenza 2023-2027, Universit{\`a} di Genova, Italy, Via Dodecaneso 35, 16146 Genova, Italy}
\affil[2]{Istituto Nazionale di Fisica Nucleare, Sezione di Genova, Via Dodecaneso 33, 16146 Genova, Italy}
\affil[3]{Institut f{\"u}r Theoretische Physik, Universit{\"a}t Leipzig, Br{\"u}derstra{\ss}e 16, 04103 Leipzig, Germany}
\affil[4]{Dipartimento di Matematica, Universit{\`a} degli Studi di Trento, Via Sommarive 14, 38123 Povo, Italy}
\affil[5]{Trento Institute for Fundamental Physics and Applications (TIFPA-INFN), Via Sommarive 14, 38123 Povo, Italy}
\affil[6]{Department of Mathematics, Faculty of Science, Kuwait University, Safat 13060, Kuwait}

\aemail{$\ast$}{edoardo.dangelo@edu.unige.it}
\aemail{$\dagger$}{mfroeb@itp.uni-leipzig.de}
\aemail{$\ddag$}{stefano.galanda@dima.unige.it}
\aemail{$\S$}{paolo.meda@unitn.it}
\aemail{$\P$}{much@itp.uni-leipzig.de}
\aemail[Corresponding author,]{$\parallel$}{kyriakos@sci.kuniv.edu.kw}

\begin{abstract}
We derive an entropy-area law for the future horizon of an observer in diamonds inside the static patch of de Sitter spacetime, taking into account the backreaction of quantum matter fields. We prove positivity and convexity of the relative entropy for coherent states using Tomita--Takesaki modular theory, from which the QNEC for diamonds follows. Furthermore, we show that the generalized entropy conjecture holds. Finally, we reveal that the local temperature which is measured by an observer at rest exhibits subleading quantum corrections with respect to the well-known cosmological horizon temperature $H/(2\pi)$.
\end{abstract}

\maketitle

\section{Introduction}
Since Bekenstein and Hawking's discovery of black hole entropy \cite{bekenstein1972,hawking1975}, the connection between entropy and geometry provided fundamental insights into the nature of quantum gravity \cite{jacobson2016}. Moreover, considerations motivated by quantum gravity fostered the study of information-theoretic quantities and new entropic measures, providing sometimes surprising results in Quantum Field Theory (QFT). One example is the introduction of entanglement entropy, as a measure of entanglement between quantum degrees of freedom in spatially separated regions, motivated by the search of a microscopic explanation of black hole entropy \cite{bombelli1986,srednicki1993,solodukhin2011}. From its introduction in the context of black hole thermodynamics, entanglement entropy found numerous applications in QFT, in particular in Conformal Field Theory in low dimensions; see e.g. \cite{holzheylarsenwilczek1994,calabresecardy2004,calabresecardy2009, wall2011,wall2017,casinihuerta2023,buenocasiniandinomoreno2023, bostelmanncadamurodelvecchio2020}. However, the entanglement entropy in QFT suffers from universal UV divergences; physically, they arise from the entanglement of modes with arbitrary high energies. Mathematically, they arise because the typical operator algebras describing observables in QFT are von Neumann factors of type III that do not admit a trace, such that a reduced density matrix does not exist \cite{buchholzfredenhagendantoni1987,borchers2000,yngvason2005,witten2018}.

A more suitable notion of entropy for the algebras of QFT is the \emph{relative entropy}, also known as \emph{Kullback--Leibler divergence} in information theory. The relative entropy measures the distinguishability between two states. It is defined for any von Neumann algebra, and so in particular it is a well-defined notion for the type III factors of QFT; in the case of Quantum Mechanics it reduces to the entanglement entropy with a vacuum-subtracted contribution. Using the Tomita--Takesaki theory of modular automorphisms \cite{tomita1967,takesaki1970}, the relative entropy can be obtained from the \emph{Araki--Uhlmann formula} \cite{araki1975,araki1976,uhlmann1977}, employing the \emph{relative modular Hamiltonian}.

In the most general setting, Tomita--Takesaki modular theory considers a von Neumann algebra $\mathfrak{A}$ acting on a Hilbert space $\mathscr{H}$, and a cyclic and separating vector $\ket{\Omega} \in \mathscr{H}$. The \emph{Tomita--Takesaki theorem} states that there exists a self-adjoint operator called \emph{modular Hamiltonian} $\mathcal{H}$, which defines an automorphism of the algebra via the \emph{modular flow} $\sigma_\tau \colon \mathfrak{A} \to \mathe^{\mathi \mathcal{H} \tau} \, \mathfrak{A} \, \mathe^{-\mathi \mathcal{H} \tau}$. Tomita--Takesaki theory thus provides a natural notion of time evolution along the modular parameter $\tau$, for which the state $\ket{\Omega}$ is thermal (KMS). Generalizing to two cyclic and separating vectors $\Omega$ and $\Phi$, the modular theory also defines a relative modular Hamiltonian $\mathcal{H}_{\Omega \vert \Phi}$. The Araki--Uhlmann formula then defines the relative entropy as~\cite{araki1976,uhlmann1977}
\begin{equation}
\label{eq:araki}
\mathcal{S}(\Omega \Vert \Phi) = - \bra{\Omega} \mathcal{H}_{\Omega \vert \Phi} \ket{\Omega} \eqend{.}
\end{equation}

Recently, relative entropy found applications in the context of semi-classical gravity, generalizing the results obtained using entanglement entropy and establishing them rigorously in QFT. For coherent states, obtained as a unitary excitation of the vacuum, the relative entropy can be computed from the \emph{modular Hamiltonian} of the vacuum alone, greatly simplifying the computations and allowing for a more direct evaluation of the entropy \cite{longo2019,casinigrillopontello2019,hollands2019,galandamuchverch2023}. It has been used to rigorously formulate and prove the Bekenstein bound \cite{bousso2003,casini2008,idaokamotosaito2013,longoxu2018}, and the Quantum Null Energy Inequality (QNEC) \cite{ceyhanfaulkner2018,ciollilongoranalloruzzi2022}, to derive the Bekenstein--Hawking formula for Schwarzschild \cite{hollands2019}, prove singularity theorems~\cite{boussoetal2022}, and to define an entropy \cite{dangelo2021} and temperature \cite{kurpiczpinamontiverch2021} for dynamical black holes.
Finally, the relative entropy between coherent states has been recently studied in de Sitter (dS) spacetime by three of the authors \cite{froebmuchpapadopoulos2023}.

De Sitter spacetime also plays an important role as testbed for quantum gravity approaches, and in particular for the \emph{generalized entropy conjecture}. The generalized entropy $\mathcal{S}_\text{gen}$ is defined as the sum of matter entropy $\mathcal{S}_\text{M}$ and the rescaled area $A_\text{dS}/(4 G_\text{N})$ of the de~Sitter horizon, and it has been conjectured that $\mathcal{S}_\text{gen}$ can only decrease when adding any type of matter~\cite{maeda1997, bousso2000, giddingsmarolf2007,banihashemijacobsonsveskovisser2023,balasubramaniannomuraugajin2023,chandrasekaranlongopeningtonwitten2023}. While the horizon entropy has been studied extensively in a holographic context~\cite{boussoengelhardt2015, nguyen2017,narayan2018,dongsilversteintorroba2018,narayan2019,genggrieningerkarch2019,ariasdiazsundell2020,geng2020,geng2021,arenashenriquezetal2022}, the conjecture itself does not make any reference to holography. It can thus be understood (and possibly proven) in a purely QFT context, taking into account the backreaction of the matter on the geometry. Here, we tackle this problem in the context of modular theory.

Concretely, we consider the relative entropy between coherent states for a massless real scalar field in dS diamonds. This is made possible by the results of Ref.~\cite{froeb2023}, where the modular Hamiltonian for these regions was derived. We show that the relative entropy is convex if we shrink the diamond in the null direction, thus proving furthermore the \emph{Quantum Null Energy Condition} (QNEC)~\cite{boussoetal2016} for \emph{finite} regions in de Sitter space.

We then associate the relative entropy to the geometric area of the cosmological horizon. In this way, we derive a conservation law for the generalized entropy, using the back-reaction of the entropy of matter fields on the classical geometry. Our result holds for the relative entropy between the vacuum and a coherent perturbation in a von Neumann algebras of type III, coupled with classical gravity through Einstein's equations. The computation can be connected with recent results based on von Neumann algebras of type II, obtained by adding to the algebra of observables for the scalar field linear, classical gravitational perturbations, coupled via the Raychaudhuri equation \cite{chandrasekaranlongopeningtonwitten2023}. However, our construction relies on the relative entropy, which is always well-defined, while the methods based on von Neumann entropies require divergent quantities at intermediate steps \cite{chandrasekaranpeningtonwitten2022,kudler-flam2023}.

Finally, we define a local notion of temperature for observers at rest inside a dS diamond. In the limit where the diamonds become large and coincide with the full static patch, we recover the well-known temperature of the cosmological horizon~\cite{figarietal1975,gibbonshawking1977}, while for finite-size diamonds we find exponentially suppressed corrections.

\section{Relative entropy}

\begin{figure}[ht]
\includegraphics{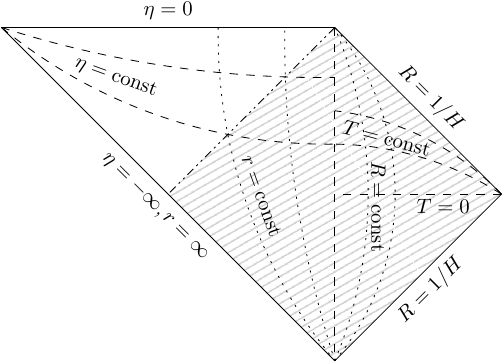}
\caption{Geometry of the patches of de~Sitter space relevant for our universe. On the left, the expanding Poincaré patch in conformally flat coordinates $(\eta,\vec{x})$ with $r = \abs{\vec{x}}$; on the right, the static patch in static coordinates $(T,\vec{X})$ with $R = \abs{\vec{X}}$. The cosmological horizon is situated at $R = H^{-1}$.}
\label{fig:desitter}
\end{figure}

The relevant part of dS is the expanding Poincaré patch with metric $g_{\mu\nu} = \mathe^{2 \omega} \eta_{\mu\nu}$ with the conformal factor $\omega = - \ln(-H \eta)$, where $\eta \in (-\infty,0)$ is the conformal time. However, not all of this is accessible to a single observer. The corresponding region (for an observer at rest at $r = 0$) is the static patch with metric
\begin{equation}
\label{eq:metric-static-coordinates}
\total s^2 = - (1 - H^2 R^2) \total T^2 + (1 - H^2 R^2)^{-1} \total R^2 + R^2 \total \Omega^2
\end{equation}
in spherical coordinates; both regions are shown in Fig.~\ref{fig:desitter}.

The dS diamonds that we consider, centered at $(\chi,\vec{0})$ and of size $\ell$, are the regions
\begin{equation}
\label{eq:desitter_diamond}
\doublecone = \left\{ (\eta,\vec{x})\colon r = \abs{\vec{x}} \in [0, \ell), \eta \in (\chi - \ell + r, \chi + \ell - r) \right\} \eqend{.}
\end{equation}
The modular Hamiltonian for conformal scalar fields $\Phi$ of dimension $\Delta$ in the conformal vacuum $\ket{\Omega}$ is given via a surface integral of the quantum stress tensor $\hat{T}_{\mu\nu}$ on a Cauchy surface $\Sigma$~\cite{froeb2023}, which we take to be the surface $\Sigma = \{ \eta = \chi \}$. It is useful to perform a conformal rescaling, writing $\Phi(f) = \phi(f_\omega)$ with the conformally rescaled test function $f_\omega(x) \equiv \mathe^{(4-\Delta) \omega(x)} f(x)$ and the flat-space massless scalar field $\phi$. The modular Hamiltonian then reads \cite{froeb2023}
\begin{equation}
\label{eq:modular-hamiltonian}
\mathcal{H}_{\chi,\ell} = \int_\Sigma \hat{T}_{\mu\nu}[\phi] \xi^\mu n^\nu \total^3 \Sigma \eqend{,}
\end{equation}
where the flat-space stress tensor $\hat{T}_{\mu\nu}[\phi]$ is contracted with the conformal Killing vector
\begin{equation}
\label{eq:conformal_killing}
\xi^\mu = \frac{\pi}{\ell} \left[ \left( \ell^2 - \tau^2 \right) \delta^\mu_0 + 2 (\tau-\eta) x^\mu - x^2 \delta^\mu_0 \right] \eqend{.}
\end{equation}

From the modular Hamiltonian we can provide an explicit formula for the relative entropy between coherent states in dS diamonds. For coherent states $\mathe^{\mathi \Phi(f)} \ket{\Omega}$, the Araki--Uhlmann formula \eqref{eq:araki} reduces to~\cite{casinigrillopontello2019,longo2019,lashkariliurajagopal2021}
\begin{equation}
\label{eq:araki-uhlmann}
\mathcal{S}\left( \Omega \Vert \mathe^{\mathi \Phi(f)} \Omega \right) = - \bra{\Omega} \mathe^{\mathi \Phi(f)} \, \mathcal{H}_{\chi,\ell} \, \mathe^{- \mathi \Phi(f)} \ket{\Omega} \eqend{.}
\end{equation}
Since $\mathcal{H}_{\chi,\ell}$ is quadratic in the fields, the above conformal rescaling and the Baker--Campbell--Haussdorf formula~\cite{achillesbonfiglioli2012} give
\begin{splitequation}
\label{eq:relative-entropy-modular-hamiltonian}
\mathcal{S}\left( \Omega \Vert \mathe^{\mathi \Phi(f)} \Omega \right) &= \frac{1}{2} \bra{\Omega} \big[ \phi(f_\omega), \big[ \phi(f_\omega), \mathcal{H}_{\chi,\ell} \big] \, \big] \ket{\Omega} \\
&= - \frac{1}{2} \iint \frac{\delta^2 \mathcal{H}_{\chi,\ell}}{\delta \phi(x) \delta \phi(y)} (\Delta f_\omega)(x) (\Delta f_\omega)(y) \total^4 x \total^4 y \eqend{,}
\end{splitequation}
where $(\Delta f_\omega)(x) \equiv \int \Delta(x,y) f_\omega(y) \total^4 y$ with the commutator function $\Delta(x,y) = - \mathi [ \phi(x), \phi(y) ]$.

Since a coherent excitation of the vacuum can also be interpreted as a classical wave, it is not surprising that we can reformulate the relative entropy as an integral over the corresponding classical stress tensor. Namely, we have
\begin{equation}
\iint\! \frac{\delta^2 \hat{T}_{\mu\nu}}{\delta \phi(x) \delta \phi(y)} (\Delta f_\omega)(x) (\Delta f_\omega)(y) \total^4 x \total^4 y = 2 T_{\mu\nu}(\Delta f_\omega)
\end{equation}
with the classical (improved) stress tensor~\cite{froeb2023}
\begin{equation}
\label{eq:SET-coherent}
T_{\mu\nu}(f) = \frac{2}{3} \partial_\mu f \partial_\nu f - \frac{1}{3} f \partial_\mu \partial_\nu f - \frac{1}{6} \eta_{\mu\nu} \partial_\rho f \partial^\rho f \eqend{,}
\end{equation}
and thus
\begin{equation}
\label{eq:relative-entropy-stress}
\mathcal{S}\left( \Omega \Vert \mathe^{\mathi \Phi(f)} \Omega \right) = \int_\Sigma T_{\mu\nu}(\Delta f_\omega) \xi^\mu n^\nu \total^3 \Sigma \eqend{.}
\end{equation}
With the Cauchy hypersurface $\Sigma = \{ \eta = \chi \}$, using the fact that $\Delta$ satisfies the Klein--Gordon equation, using the result~\eqref{eq:SET-coherent} and integrating spatial derivatives by parts, we find that the relative entropy is given by the relatively simple expression
\begin{equation}
\label{eq:entropy-diamond}
\mathcal{S}\left( \Omega \Vert \mathe^{\mathi \Phi(f)} \Omega \right) = \frac{\pi}{2 \ell} \int_{\eta = \chi} \Big[ \left( \ell^2 - \vec{x}^2 \right) ( \partial_\eta (\Delta f_\omega) \partial_\eta (\Delta f_\omega) + \partial_i (\Delta f_\omega) \partial^i (\Delta f_\omega) ) + 2 (\Delta f_\omega)^2 \Big] \total^3 \vec{x} \eqend{.}
\end{equation}
This expression of the relative entropy is manifestly positive, as required from general considerations \cite{hollandssanders2017}.

\section{Half-sided modular inclusions and QNEC}

Apart from positivity, the relative entropy given by the Araki--Uhlmann formula~\eqref{eq:araki-uhlmann} satisfies many other important properties such as monotonicity under completely positive maps and joint convexity in both arguments \cite{hollandssanders2017}. It also provides a basis to rigorously formulate and prove entropy inequalities in QFT. In particular, it is known that the relative entropy between coherent states is convex under half-sided modular inclusions, which is a sufficient condition to prove the QNEC on any half-invariant \emph{wedge} on a globally hyperbolic spacetime \cite{ciollilongoranalloruzzi2022}.

\begin{figure}[ht]
\includegraphics{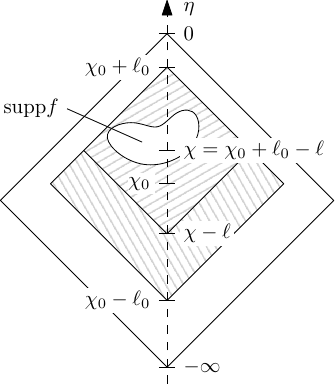}
\caption{The diamonds relevant for the half-sided modular inclusion. Note that their future boundary coincides, and that we take a test function $f$ whose support is inside the smaller diamond.}
\label{fig:diamonds-inclusion}
\end{figure}

Here, we prove that the relative entropy is convex also under inclusions for dS \emph{diamonds}, by explicit computation. In this context, half-sided modular inclusions are realized geometrically by shrinking diamonds towards the future, keeping the future tip fixed. For illustration, consider a diamond of size $\ell_0$ with center at $(\chi_0,\vec{0})$, and a diamond of size $\ell < \ell_0$ with center at $(\chi_0 + \ell_0 - \ell,\vec{0})$, see Fig.~\ref{fig:diamonds-inclusion}. The modular flow for the modular Hamiltonian of the bigger diamond then satisfies
\begin{equation}
\mathe^{\mathi \tau \mathcal{H}_{\chi,\ell_0}} \, \mathfrak{A}_\text{sub} \, \mathe^{- \mathi \tau \mathcal{H}_{\chi,\ell_0}} \subset \mathfrak{A}_\text{sub} \eqend{,} \quad \tau \geq 0 \eqend{,}
\end{equation}
where $\mathfrak{A}_\text{sub}$ is the subalgebra of fields with support inside the smaller diamond. This is exactly the condition for a half-sided modular inclusion \cite{borchers2000,ciollilongoranalloruzzi2022}.

To show that the entropy is convex under these inclusions, we take a Cauchy surface at $\eta = \chi = \chi_0 + \ell_0 - \ell$, and assume that $\Delta f_\omega$ has compact support on this Cauchy surface. Then the relative entropy is given by Eq.~\eqref{eq:entropy-diamond} with $\chi = \chi_0 + \ell_0 - \ell$, and a short computation results in
\begin{equation}
\partial_\ell^2 \mathcal{S}\left( \Omega \Vert \mathe^{\mathi \Phi(f)} \Omega \right) = - \frac{2}{\ell} \partial_\ell \mathcal{S}\left( \Omega \Vert \mathe^{\mathi \Phi(f)} \Omega \right) \eqend{.}
\end{equation}
This expression is easily integrated, resulting in
\begin{equation}
\mathcal{S}\left( \Omega \Vert \mathe^{\mathi \Phi(f)} \Omega \right) = \frac{\mathcal{S}_1}{\ell} + \mathcal{S}_2
\end{equation}
with two expressions $\mathcal{S}_i$ independent of $\ell$.\footnote{This form also follows from the explicit expression of the action of the modular operator $\mathcal{H}_{\chi,\ell}$ in static coordinates~\cite{froeb2023}.}

Since the relative entropy is positive for any $\ell$, it must be that both $\mathcal{S}_1, \mathcal{S}_2 > 0$. Thus it follows that
\begin{equation}
\label{eq:entropy-convexity}
\partial_\ell^2 \mathcal{S}\left( \Omega \Vert \mathe^{\mathi \Phi(f)} \Omega \right) \geq 0 \eqend{,}
\end{equation}
proving convexity for the relative entropy of dS diamonds. Since the geometric transformation is a shrinking of the dS diamond in the null direction, formula~\eqref{eq:entropy-convexity} is sufficient to prove the QNEC \cite{ciollilongoranalloruzzi2022} for these regions.

\section{Entropy-area law from the modular flow and backreaction}

It has been known for a long time that it is possible to associate an entropy to cosmological horizons \cite{gibbonshawking1977}, similarly to the Bekenstein-Hawking entropy of black holes \cite{hawking1975}. We now show that the backreaction of the matter perturbation on the classical geometry leads to a change in the cosmological horizon area consistent with the generalized entropy conjecture for dS.

\begin{figure}[ht]
\includegraphics{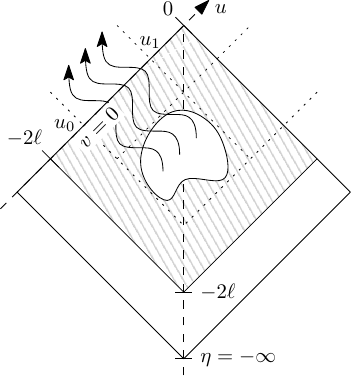}
\caption{De Sitter diamond of size $2\ell$ whose future coincides with the cosmological horizon at $v = 0$. Since we are considering massless fields in even dimensions, the horizon is perturbed only for some null time $u \in [u_0,u_1] \subset [-2\ell,0]$.}
\label{fig:diamonds-area}
\end{figure}

Analogously to the computations for Schwarzschild \cite{hollands2019} and dynamical black holes \cite{dangelo2021}, we employ the Raychaudhuri equation to relate the backreaction of the relative entropy to the variation of the horizon area. Consider the diamond \eqref{eq:desitter_diamond} centered at $\chi = - \ell$, and introduce null coordinates $u = \eta-r$, $v = \eta+r$. The metric becomes
\begin{equation}
\total s^2 = \frac{4}{H^2 (u+v)^2} \left[ - \total u \total v + \frac{(v-u)^2}{4} \total \Omega^2 \right] \eqend{,}
\end{equation}
and the future cosmological horizon is the surface $v = 0$, $u \in [-2\ell,0]$, see Fig.~\ref{fig:diamonds-area}. The horizon is described by intrinsic coordinates $u,\theta,\phi$, and the normal vector $n_\mu = \partial_\mu v = \delta_\mu^v$, or $n^\mu = - 2 \delta^\mu_u$. After a conformal rescaling, the induced cross-sectional metric of the horizon is the one of the unit two-sphere with integration measure $\total \Omega = \sin \theta \total \theta \total \phi$, and the conformal Killing vector~\eqref{eq:conformal_killing} inducing the modular flow reads
\begin{equation}
\xi^\mu \partial_\mu = - \frac{\pi}{\ell} \big[ u (u+2\ell) \partial_u + v (v+2\ell) \partial_v \big] \eqend{.}
\end{equation}

Since the formula for the relative entropy is independent on the choice of the Cauchy surface $\Sigma$ on which the surface integral is computed \cite{froeb2023}, we can consider the limit in which $\Sigma$ coincides with the cosmological horizon. Eq.~\eqref{eq:relative-entropy-stress} for the relative entropy, evaluated on the horizon, then simplifies to
\begin{equation}
\label{eq:relative-entropy-stress-horizon}
\mathcal{S}\left( \Omega \Vert \mathe^{\mathi \Phi(f)} \Omega \right) = - \frac{2 \pi}{\ell} \int_{-2\ell}^0 \int T_{uu} \big\rvert_{v = 0} u (u+2\ell) \total \Omega \total u \eqend{.}
\end{equation}

To determine the change of the horizon area to leading order in perturbations, we compute the backreaction of the matter on the geometry analogously to the case of black holes \cite{hollandsishibashi2019}. The leading correction to the metric under the influence of the stress tensor $T_{\mu\nu}(\Delta f_\omega)$~\eqref{eq:SET-coherent} is quadratic in $f_\omega$. Hence, the leading-order correction to the Raychaudhuri equation, in null coordinates, determines the backreaction on the background geometry,
\begin{equation}
\left. \frac{\total \delta \Theta}{\total u} \right\rvert_{v = 0} = - 32 \pi G_\text{N} T_{uu} \eqend{,}
\end{equation}
where $\delta \Theta$ denotes the geodesic expansion of the cosmological horizon due to the coherent perturbation. Both the shear and vorticity tensors, as well as the contribution quadratic in $\Theta$, do not contribute to the right-hand side of the Raychaudhuri equation when evaluated at second order in perturbations \cite{hollandsishibashi2019}.

Multiplying the Raychaudhuri equation with $u(u+2\ell)$ and integrating over $u$ and $\Omega$, Eq.~\eqref{eq:relative-entropy-stress} can be written as
\begin{equation}
\mathcal{S}\left( \Omega \Vert \mathe^{\mathi \Phi(f)} \Omega \right) =  \int_{-2\ell}^0 \frac{u (u+2\ell)}{16 G_\text{N} \ell} \int \left. \frac{\total \delta \Theta}{\total u} \right\rvert_{v = 0} \total \Omega \total u \eqend{.}
\end{equation}
Since on the background we have $\total \Theta/\total u = 0$, to leading order in perturbations we can replace $\delta \Theta$ by the complete geodesic expansion $\Theta + \delta \Theta = \tilde{\Theta}$. In the same way, $\total \Omega$ is substituted by the complete line element $\total \tilde{\Omega} = \sqrt{\gamma} \total \theta \total \phi$. Then we use the definition of the expansion as the logarithmic derivative of the cross-sectional area of the geodesic congruence \cite{wald1984}, $\tilde{\Theta} \equiv \total (\ln \sqrt{\gamma})/\total u$, and integrate by parts. This results in
\begin{equation}
\label{eq:entropy_area_average}
\mathcal{S}\left( \Omega \Vert \mathe^{\mathi \Phi(f)} \Omega \right) = \frac{1}{4 G_\text{N}} \llangle A_\text{H} \rrangle - \frac{A_\text{H}(0) + A_\text{H}(-2\ell)}{8 G_\text{N}} \eqend{,}
\end{equation}
where the cross-sectional area of the perturbed cosmological horizon is given by
\begin{equation}
A_\text{H}(u) \equiv \int \sqrt{\gamma(u)} \total \theta \total \phi \eqend{,}
\end{equation}
and we defined the null average
\begin{equation}
\llangle A \rrangle = \frac{1}{2 \ell} \int_{-2\ell}^0 A(u) \total u \eqend{.}
\end{equation}

Since the background area of the cosmological horizon is constant, in Eq.~\eqref{eq:entropy_area_average} only the perturbations contribute. Moreover, the fields are massless and conformally coupled and thus propagate along null geodesics in the conformally flat de~Sitter spacetime~\cite{friedlander1975}, see Fig.~\ref{fig:diamonds-area}. It follows that the area perturbations at future infinity (i.e., at $u = v = 0$) and at the rim of the diamond (i.e., at $u = - 2 \ell$, $v = 0$) vanish, $\delta A_\text{H}(0) = 0 = \delta A_\text{H}(-2\ell)$, and thus only the average of the perturbation of the area remains:
\begin{equation}
\mathcal{S}\left( \Omega \Vert \mathe^{\mathi \Phi(f)} \Omega \right) = \frac{1}{4 G_\text{N}} \llangle[\big] \delta A_\text{H} \rrangle[\big] \eqend{.}
\end{equation}
It follows that to leading order in matter perturbations, the perturbation of generalized entropy $\delta \mathcal{S}_\text{gen} = - \mathcal{S}\left( \Omega \Vert \mathe^{\mathi \Phi(f)} \Omega \right) + \frac{1}{4 G_\text{N}} \llangle \delta A_\text{H} \rrangle$ vanishes for dS diamonds. We note that the minus sign in the generalized entropy, which gave raise to some subtle interpretations \cite{banihashemijacobsonsveskovisser2023}, naturally arises in this context. Hence the generalized entropy conjecture, according to which $\mathcal{S}_\text{gen}$ does not increase when perturbing dS, holds in our case, and we expect that it decreases when we take into account also the backreaction of quantized metric perturbations.

\section{Local temperature}

Finally, the thermodynamic relation between entropy and energy provides a notion of local temperature. Consider an observer with proper time $t$ and four-velocity $v^\mu = - g^{\mu\nu} \partial_\nu t$ inside a diamond, who crosses some Cauchy surface $\Sigma$ normally such that $v_\mu \big\rvert_\Sigma = n_\mu$ with the normal vector $n_\mu$. The relative entropy measured by this observer is given by Eq.~\eqref{eq:relative-entropy-stress}, and the entropy density $s$ by the integrand in this formula. On the other hand, the observer measures the energy density $e \equiv v^\mu v^\nu T_{\mu\nu}$, and the first law of thermodynamics results in a local temperature: $\delta s = \beta \delta e$. Comparing the two expressions for $s$ and $e$, we find that $\beta = \partial t/\partial (-\tau)$ where $\tau$ is the parameter of the modular flow generated by the modular Hamiltonian~\eqref{eq:modular-hamiltonian}, i.e., $\xi^\mu \partial_\mu = \partial_\tau$ with the conformal Killing vector $\xi^\mu$~\eqref{eq:conformal_killing}. This is, in fact, the temperature arising from the thermal time hypothesis \cite{connesrovelli1994,martinettirovelli2003,longomartinettirehren2010}.

To obtain an explicit expression for the (inverse) temperature $\beta$, we switch to static coordinates. In those, the diamonds of Fig.~\ref{fig:diamonds-area} read
\begin{equation}
\doublecone = \left\{ (T,\vec{X})\colon H R < 1, H T > - \ln\left( 2 H \ell \sqrt{\frac{1 - H R}{1 + H R}} \right) \right\}
\end{equation}
with $R = \abs{\vec{X}}$. An observer at rest passing through the point $(T,\vec{0})$ at $\tau = 0$ follows the trajectory $(T,\vec{X})(\tau) = (T_\tau,\vec{0})$ with~\cite{froeb2023}
\begin{equation}
\label{eq:desitter_doublecone_flow_coords_static}
T_\tau = T_\text{min} + \frac{1}{H} \ln\left[ 1 + \mathe^{- 2 \pi \tau} \left( 2 \mathe^{H T} H \ell - 1 \right) \right] \eqend{.}
\end{equation}
Since
\begin{equation}
\lim_{\tau \to \infty} T_\tau = T_\text{min} \equiv - \frac{1}{H} \ln(2 H \ell)
\end{equation}
is the static time of the lower tip of the diamond, and $\lim_{\tau \to -\infty} T_\tau = \infty$ the one of the upper tip, the observer traverses the whole diamond.

Computing $\beta = \partial T_\tau/\partial (-\tau)$, we obtain
\begin{equation}
\beta = \frac{2 \pi}{H} \frac{2 \mathe^{H T} H \ell - 1}{2 \mathe^{H T} H \ell - 1 + \mathe^{2 \pi \tau}} = \frac{2 \pi}{H} \left[ 1 - \mathe^{- H ( T_\tau - T_\text{min} )} \right] \eqend{.}
\end{equation}
The first term recovers the well-known temperature associated with cosmological horizons~\cite{gibbonshawking1977}, but the second one is a subleading correction, which in fact decays exponentially fast with time. However, in the limit $\ell \to \infty$ where the diamond becomes large and coincides with the full static patch, we have $T_\text{min} \to -\infty$ and recover $\beta \to 2 \pi/H$ for all times.

\section{Conclusion and outlook}

We determined the (Araki--Uhlmann) relative entropy for coherent excitations of the vacuum inside dS diamonds. We have shown that the relative entropy is convex when shrinking the diamond in a null direction, which proves the QNEC~\cite{boussoetal2016} for diamonds in dS. Considering then the backreaction of the coherent excitation on the geometry via the Raychaudhuri equation, we proved that the change in the area of the future horizon exactly compensates the relative entropy, such that the generalized entropy remains constant. Finally, we showed that an observer inside a diamond measures a local temperature which is different from the dS temperature $H/(2\pi)$, but that the corrections decay exponentially fast.

The viewpoint assumed in this work provides a framework for more general backreaction problems, that can be explored in future works. In particular, it is important to study relative entropy for quantum gravitational perturbations, and the induced perturbations of the cosmological horizon. Including these corrections should decrease the horizon area, verifying thus the generalized entropy conjecture. To address the issue of gauge invariance, it will be necessary to employ relational observables~\cite{dittrich2006,gieselthiemann2015,brunfredhackpinrej2016,froeblima2023}.

\section*{Acknowledgment}
\ 
We are grateful to Nicola Pinamonti and Rainer Verch for useful discussions. M.B.F. thanks the University of Genova, and E.D., S.G., and P.M. thank Daniela Cadamuro and the ITP Leipzig for their kind hospitality.  M.B.F. is supported by the Deutsche Forschungsgemeinschaft (DFG, German Research Foundation) --- project no. 396692871 within the Emmy Noether grant CA1850/1-1. E.D. is supported by a PhD scholarship of the University of Genoa and by the project GNFM-INdAM Progetto Giovani \emph{Non-linear sigma models and the Lorentzian Wetterich equation}, CUP\textunderscore E53C22001930001. E.D., S.G., and P.M. are grateful for the support of the National Group of Mathematical Physics (GNFM-INdAM). The research performed by E.D. and S.G. was supported in part by the MIUR Excellence Department Project 2023-2027 awarded to the Dipartimento di Matematica of the University of Genova, CUP\textunderscore D33C23001110001.

\bibliographystyle{JHEP-for-ptep}
\bibliography{literature}

\end{document}